\documentclass{ifacconf}
\usepackage[T1]{fontenc}
\usepackage{amsmath}
\usepackage{psfrag}
\usepackage{amssymb}
\usepackage{color}
\usepackage{graphicx}      
\usepackage{natbib}        
\begin{document}
\begin{frontmatter}

\title{Nuclear Norm Subspace Identification (N2SID) for short data batches\thanksref{footnoteinfo}} 

\thanks[footnoteinfo]{Part of the research was done while the first author
 was a Visiting Professor at the
Division of Automatic Control, Department of Electrical Engineering, 
Link\"oping University, Sweden.
This work was partially supported by the European Research
Council Advanced Grant Agreement No. 339681. Corresponding Author: 
{\tt\small  m.verhaegen@tudelft.nl.}}

\author[First]{Michel Verhaegen} 
\author[Second]{Anders Hansson} 

\address[First]{Delft Center for Systems and
  Control\\ Delft University\\Delft, The Netherlands}
\address[Second]{Division of Automatic Control\\
Link\"oping University\\ Link\"oping, Sweden}

\begin{abstract}                
Subspace identification is revisited in the scope of nuclear norm
minimization methods. It is shown that essential structural knowledge
about the unknown data matrices in the \emph{data equation} that relates
Hankel matrices constructed from input and output data can be used in
the first step of the numerical solution presented. The structural
knowledge comprises the low rank property of a matrix that is the
product of the extended observability matrix and the state sequence
and the Toeplitz structure of the matrix of Markov parameters (of the
system in innovation form). The new subspace identification method is
referred to as the N2SID (\emph{twice} the N of Nuclear Norm and SID
for Subspace IDentification) method. In addition to include key
structural knowledge in the solution it integrates the subspace
calculation with minimization of a classical prediction error cost
function. The nuclear norm relaxation enables us to perform such
integration while preserving \emph{convexity}. The advantages of N2SID
are demonstrated in a numerical \emph{open- and closed-loop}
simulation study. Here a comparison is made with another widely used
SID method, i.e. N4SID. The comparison focusses on the identification
with short data batches, i.e. where the number of measurements is a
small multiple of the system order. 
\end{abstract}


\end{frontmatter}

\section{Introduction}

System identification is a key problem in a large number of scientific
areas. Generally there are two families of system identification
methods: (1) prediction error methods and (2) subspace methods,
\cite{Ljung:99,VerBook}. Either
of these approaches can be treated in the time or frequency domain, and for the
sake of simplicity we restrict ourselves to the time domain in this paper.

The central theme in prediction error methods is to parametrize the
predictor (observer) to generate an estimate of the output and then
formulate an optimization problem to minimize a (weighted) cost
function defined on the difference between the measured output and the
observer predicted output. This cost function generally is a sample
average (for the finite data length case) of the trace of the
covariance matrix of the prediction error. Though the prediction error
framework provides a vast amount of insights in studying and analyzing
the estimated predictor, its main drawback is the non-convexity for
general multi-variable state space models in innovation form, as
considered in this paper. The lack of convexity can result in that the
optimization method get stuck in a local minimum, and thereby
complicating the analysis of the numerical results, such as e.g. the
difficulty to distinguish between a bad model estimate due to a local
minimum or due to a bad model parametrization. This parametrization
needs to be given before starting the parameter optimization problem,
and thus the use of the approach can be quite complex and labor intensive
for the non-expert user. However, the latter fact has been greatly
relieved by computer added software packages such as in
\cite{Ljungtb:07}. 

Motivated by the drawbacks of prediction error methods, the goal with
subspace identification methods is to derive \emph{approximate}
models rather than models that are ``optimal'' with respect to a
chosen cost function. The approximation is based on linear algebra
transformations and factorizations with structured Hankel matrices
constructed form the input-output data. All existing subspace
identification methods aim to derive a low rank matrix
from which key subspaces, hence the name subspace identification, are
derived. The low rank approximation is in general done using a
\emph{singular value decomposition} (SVD). Recently a new family of subspace
identification methods was presented that use the \emph{nuclear norm}
instead of a SVD in order to improve the low rank
approximation,
\cite{Liu+van09,LiV:09,MoF:10,faz+pon+sun+tse12,HLV:12,liu+han+van13}.

The drawback of Subspace Identification (SID) is two-fold. First general
subspace identification methods lack 
an optimization criterion in the calculation of the predictor. This
drawback is relaxed in a number of recent nuclear norm based SID
methods that regularize the low rank approximation problem with a 
 sample average of the trace of the covariance matrix of the
prediction error. The second drawback is that the low rank
approximation is not performed in the first step of the algorithm. As
a consequence, this low rank decomposition does not operate with the original
input-output data but with approximate processed data. The latter
approximation destroys the low rank property and 
especially for short data batches,
since most of the SID schemes only provide consistent estimates. 

As it is known that exploiting structure (that is present in the
model) is beneficial, especially when working with short data batches,
we review the derivation of subspace identification in order to be
able to deal with two key structure properties in the data equation used
by SID methods. The data equation is a relationship between the Hankel
matrices constructed from the input and output data, respectively. The
key structural properties are the low rank property and the Toeplitz
structure of unknown model dependent matrices in the data
equation. The key  in the derivation of the new scheme is that both these 
structural properties are invoked in the first step of the
algorithm. The algorithm is abbreviated by N2SID, standing form Nuclear
Norm (\emph{two} times N, i.e. N2) Subspace Identification. 

The paper is organized as follows. In Section 2 the identification
problem for identifying a multivariable state space model in  a
subspace context while taking a prediction error cost function into
consideration is presented. The problem formulation does however not require a
parametrization of the system matrices of the state space model as is
classically done in prediction error methods for parametric
input-output models, \cite{Ljung:99}. The key data equation in the
analysis and description of subspace identification methods is
presented in Section 3. Here we also highlight the important
structural properties in the submatrices of this equation, the
low rank property of the matrix that is the product of the extended
observability matrix and the state sequence of the underlying state
space model (given in innovation form) and the (lower triangular)
Toeplitz structure of the model Markov parameters. A convex relaxation is
presented in Section 4 to take these structural constraints into
consideration. Some preliminary results of the performances are
illustrated in Section 5 in a comparison study of N2SID with N4SID for
both open and closed-loop identification problems with short data
batches. Here we consider the identification of second order systems
with just $50$ data points. Finally, we end this paper with some
concluding remarks.

\section{The Subspace Identification Problem}
\label{sec:S2IDE}

In system identification a challenging problem is to identify Linear
Time Invariant (LTI) systems with multiple inputs and multiple outputs
using short length data sequences. Taking process and measurement
noise into consideration, the model of the LTI system can be written in
the following innovation form:
\begin{equation}
\label{ssmI}
\left\{  \begin{array}{rcl}
x(k+1) & = & A x(k) + B u(k) + K e(k) \\
y(k) & = & C x(k) + D u(k) + e(k)
\end{array} \right.
 \end{equation}
with $x(k) \in \mathbb{R}^n, y(k) \in \mathbb{R}^p, u(k) \in
\mathbb{R}^m$ and $e(k)$ a zero-mean white noise sequences with
covariance matrix $R_e$. 

We will consider a possible open or closed loop scenario in which the
input and output data sequences  are generated. Also the case of
output only identification is considered, making the approach outlined
in  this paper a general and new framework to identify linear
dynamical systems. 

The problem considered can be formulated as follows:
\begin{quote}
\emph{
Given the input-ouput (i/o) data batches $\{u(k),y(k)\}_{k=1}^N$ that
are generated with a system belonging to the class of LTI systems as
represented by \eqref{ssmI}, the problem is using short length i/o data
(with $N$ larger but of the order of $n$) to estimate the system
order, denoted by $\hat{n}$, and to determine approximate system
matrices $(\hat{A}_T,\hat{B}_T,\hat{C}_T,\hat{D},\hat{K}_T)$ that define the observer:
 \begin{equation}
\label{ssmOT}
\left\{  \begin{array}{rcl}
 \hat{x}_T(k+1) & = & \hat{A}_T \hat{x}_T(k) + \hat{B}_T u_v(k) + \\
 & &  \hat{K}_T \Big( y_v(k) - \hat{C}_T \hat{x}_T \Big)  \\
\hat{y}_v(k) & = & \hat{C}_T \hat{x}_T(k) + \hat{D} u_v(k) 
\end{array} \right.
 \end{equation}
with $\hat{x}_T(k) \in \mathbb{R}^{\hat{n}}$, 
such that the approximated output $\hat{y}_v(k)$ is close to the
measured output $y_v(k)$ of the validation pair $\{ u_v(k),y_v(k)
\}_{k=1}^{N_v}$ as expressed by a small value of the cost function,
\begin{equation}
 \label{CostF}
 \frac{1}{N_v} \sum_{k=1}^{N_v} \| y_v(k) - \hat{y}_v(k) \|_2^2
\end{equation}
}
\end{quote}
The quantitative notions like ``small'' and ``approximate'' will be
made more precise in the new N2SID solution toward this problem. It
should be remarked that contrary to many earlier Subspace
Identification (SID) methods, the present problem formulation
explicitly takes a prediction error cost function like \eqref{CostF} into
consideration. 

A key starting point in the formulation of subspace methods is the
relation between (structured) Hankel matrices constructed from the i/o
data. This relationship  will as defined in \cite{VerBook} be
the Data Equation. It will be presented in the next
section. Here we will also highlight briefly how existing subspace
methods have missed to take important structural information about the
matrices in this equation into account from the first steps of the
solution. N2SID will overcome these shortcomings. Taking these
structural properties into consideration makes N2SID attractive
especially for identifying MIMO LTI models when only having short data
batches. 

\section{The Data Equation and its structure}
 \label{s_dataeq}
Let the LTI model \eqref{ssmI} be represented in its so-called
observer form:
\begin{equation}
\label{ssmO}
\left\{  \begin{array}{rcl}
x(k+1) & = & (A - KC) x(k) + (B - KD) u(k) + K y(k) \\
y(k) & = & C x(k) + D u(k) + e(k)
\end{array} \right.
 \end{equation}
We will denote this model compactly as:
\begin{equation}
\label{ssmOc}
\left\{  \begin{array}{rcl}
x(k+1) & = & \mathcal{A} x(k) + \mathcal{B} u(k) + K y(k) \\
y(k) & = & C x(k) + D u(k) + e(k)
\end{array} \right.
 \end{equation}
with $\mathcal{A}$ the observer system matrix $(A-KC)$ and
$\mathcal{B}$ equal to $(B-KD)$. Though this property will not be used
in the sequel, the matrix $\mathcal{A}$ can be assumed to be
asymptotically stable.

For the construction of the data equation, we store the measured i/o
data in (block-) Hankel matrices. For fixed $N$ assumed to be larger
then the order $n$ of the underlying system, the definition of the
number of block-rows fully defines the size of these Hankel
matrices. Let this dimensioning parameter be denoted by $s$, and let
$s > n$, then the
Hankel matrix of the input (similarly for the output) is defined as:
\begin{equation}
 \label{defHaki}
U_s = \left[ \begin{matrix} u(1) & u(2) & \cdots & u(N-s+1) \\ u(2) &
    u(3) &   & \vdots \\ \vdots &  & \ddots &  \\ u(s) & u(s+1) &
    \cdots &  u(N) \end{matrix} \right]
\end{equation}
The Hankel matrix defined from the output $y(k)$ and the innovation
$e(k)$ are denoted by $Y_s$ and $E_s$, respectively. The relationship between
these Hankel matrices, that readily follows from the linear model
equations in \eqref{ssmOc}, require the definition of the following
\emph{structured} matrices. First we define the (extended) observability matrix
$\mathcal{O}_s$. 
\begin{equation}
 \label{Osv}
\mathcal{O}_s^T = \left[ \begin{matrix} C^T & \mathcal{A}^T C^T & \cdots &
    \mathcal{A^T}^{s-1} C^T \end{matrix} \right] 
\end{equation}
Second, we define a Toeplitz matrix from the quadruple of systems
matrices $\{\mathcal{A},\mathcal{B},C,D\}$ as:
\begin{equation}
 \label{Toepl}
T_{u,s} = \left[ \begin{matrix} D & 0 & \cdots & 0 \\ C
    \mathcal{B} & D &   & 0 \\ \vdots &  & \ddots & \\
C \mathcal{A}^{s-2} \mathcal{B} &  & \cdots & D  \end{matrix} \right] 
\end{equation}
and in the same we define a Toeplitz matrix $T_{y,s}$ from the
quadruple $\{\mathcal{A},K,C,0\}$. Finally, let the state sequence be
  stored as:
\begin{equation}
 \label{state}
X = \left[ \begin{matrix} x(1) & x(2) & \cdots & x(N-s+1) \end{matrix}
\right] 
\end{equation}
Then the data equation compactly reads:
\begin{equation}
 \label{DataEq}
Y_s = \mathcal{O}_s X + T_{u,s} U_s + T_{y,s} Y_s + E_s
\end{equation}
This equation is a simple linear matrix equation that highlights the
challenges in subspace identification, which is to approximate from the
given Hankel matrices $Y_s$ and $U_s$ the column space of the
observability matrix and/or that of the state sequence. 

The equation is highly structured. For the sake of brevity in this
paper we focus on the following two key structural properties:
\begin{enumerate}
 \item The matrix product $\mathcal{O}_s X$ is \emph{low rank} since $s >
   n$. 
 \item The matrices $T_{u,s}$ and $T_{y,s}$ are (block-) Toeplitz.
\end{enumerate}
In all existing subspace identification methods, these two key
structural matrix properties are \emph{not used} in the first step of the
algorithmic solution. Some pre-processing step of the data (Hankel)
matrices is usually performed, followed by a low
rank factorization. To further illustrate this point, consider the
approach in \cite{Chiuso} as an example method. Then the first step
consists in the estimation of a high order ARX model that is defined
by the last (block-) row of the data equation \eqref{DataEq}
\emph{neglecting} the term 
$C \mathcal{A}^{s-1}X$. This is based on 
the fact that $\mathcal{A}$ is asymptotically stable and the
assumption that $s$ is
chosen large enough. Then in a later step the estimated ARX parameters are
used to define a matrix that asymptotically (both in terms of $s$ and
$N$) is of low rank $n$. Other recent methods that make use of the
application of the nuclear norm, try to enforce the low rank property
to an already pre-processed matrix. For example in \cite{liu+han+van13}
the case of measurement noise was considered only, and in a
pre-processing step the  input Hankel matrix $U_s$ is annihilated from
the data equation by an orthogonal projection. 

Though statistical consistency generally holds for the existing
subspace identification schemes, they refrain from exploiting key
structural matrix properties in the data equation in the first step of
the algorithm. Such structural information may  be key when dealing
with short data batches. Therefor in the next section we will revise
subspace identification and formulate the new N2SID approach. 

\section{N2SID}

 \subsection{Pareto optimal Subspace Identification}

From the data equation \eqref{DataEq} it follows that the minimum variance
prediction of the output equals $\hat{y}(k) = y(k) - e(k)$. Let the
Hankel matrix $\hat{Y}_s$ be defined from this sequence $\hat{y}(k)$
as we defined $Y_s$ from $y(k)$. Then the data equation becomes:
\begin{equation}
 \label{DataEqPred}
\hat{Y}_s = \mathcal{O}_s X + T_{u,s} U_s + T_{y,s} Y_s 
\end{equation}
Let $\mathcal{T}_{p,m}$ denote the class of lower triangular (block-) Toeplitz
matrices with block entries $p \times m$ matrices and let $\mathcal{H}_p$ denote the class
of (block-) Hankel matrices with block entries of $p$ column
vectors. Then the two
key structural properties listed in Section~\ref{s_dataeq} are taken
into account in the following problem formulation:
\begin{equation}
 \label{NPhard}
\min_{\hat{Y}_s \in \mathcal{H}_p, T_{u,s} \in  \mathcal{T}_{p,m},
  T_{y,s} \in \mathcal{T}_{p,p} } \mbox{\rm rank} \Big( \hat{Y} -
T_{u,s} U_s - T_{y,s} Y_s \Big) 
\end{equation}
and $\min \mathbb{E} [ \Big( y(k) - \hat{y}(k) \Big) \Big(
y(k) - \hat{y}(k) \Big)^T ]$, where $\mathbb{E}$ denotes the
expectation operator. This optimization problem seeks for the Pareto
optimal solution with respect to the two cost functions $\mbox{\rm rank} \Big( \hat{Y} -
T_{u,s} U_s - T_{y,s} Y_s \Big) $ and $\mathbb{E} [ \Big( y(k) - \hat{y}(k) \Big) \Big(
y(k) - \hat{y}(k) \Big)^T ]$. This optimization is 
however not tractable. For
that purpose we will develop in the next subsection a \emph{convex
relaxation}. This will make it possible to obtain all Pareto optimal 
solutions using scalarization. 
 \subsection{A convex relaxation}

A convex relaxation of the NP hard problem formulation in
\eqref{NPhard} will now be developed. The
original problem is reformulated in two ways. First the 
rank$(\cdot)$ operator is substituted by the nuclear norm. The nuclear
norm of a matrix $X$ denoted by $\| X \|_\star$ is defined as the
sum of the singular values of the matrix $X$. It is also known
as the trace norm, the Ky Fan norm or the Schatten
norm \cite{Vandenberghe}. This is known to be a good approximation,
\cite{FHB:01,Faz:02}.
Second the minimum variance criterion is
substituted by the following sample average of the trace of the
covariance matrix $\mathbb{E} [ \Big( y(k) - \hat{y}(k) \Big) \Big(
y(k) - \hat{y}(k) \Big)^T ]$:
\[
 \frac{1}{N} \sum_{k=1}^N \| y(k) - \hat{y}(k) \|_2^2
\]
By introducing a scalarization, or regularization, 
parameter $\lambda\in[0,\infty)$ 
all Pareto optimal
solutions of the convex
reformulation of the N2SID problem can be formulated in {\bf one
  line}. 
\begin{equation}
 \label{N2SID}
\left. \begin{array}{l}
\min_{\hat{Y}_s \in \mathcal{H}_p, T_{u,s} \in  \mathcal{T}_{p,m},
  T_{y,s} \in \mathcal{T}_{p,p} } \|  \hat{Y} -
T_{u,s} U_s - T_{y,s} Y_s \|_\star + \\[10pt] \quad 
   \frac{\lambda}{N} \sum_{k=1}^N \| y(k) -
   \hat{y}(k) \|_2^2 \end{array} \right. 
\end{equation}
It is well-known that this problem can be recast as a semi-definite 
programming problem, \cite{FHB:01,Faz:02}, 
and hence it can be efficiently solved with standard 
solvers. We have used the modeling language YALMIP, \cite{Lofberg:01}, 
to perform
experiments, results of which we will present next. 

The method encompasses in a straightforward manner the identification
problems with output data only. In that case the convex relaxed
problem formulation reads:
 \begin{equation}
 \label{N2SIDoutputonly}
\left. \begin{array}{l}
\min_{\hat{Y}_s \in \mathcal{H}_p, 
  T_{y,s} \in \mathcal{T}_{p,p} } \|  \hat{Y} -
T_{y,s} Y_s \|_\star + \\[10pt] \quad 
   \frac{\lambda}{N} \sum_{k=1}^N \| y(k) -
   \hat{y}(k) \|_2^2 \end{array} \right. 
\end{equation}

\section{Illustrative Examples}

\subsection{Open-loop experiment}

In this section we report results on numerical experiments in Matlab. 
In the first example we consider 100  second order single-input single-output
systems as in (\ref{ssmI})
randomly generated with the command {\tt drss}. The matrix $K$ has been 
generated with the command {\tt randn} which gives a matrix of elements 
drawn from a standardized normal density function. Any model for which 
the absolute value of the largest eigenvalue of the system matrix $A$
is larger than 0.99 has been discarded. Data for system identification has been 
generated for each model and with time horizon $N=50$. The noise 
$e(k)$ is white and drawn from a normal density function with standard 
deviation equal to 0.2. The input signal $u(k)$ is a sequence of 
$\pm 1$ obtained by taking the sign of a vector of values obtained from 
a standardized normal density function. The initial value is obtained
from a normal density function with standard deviation 5, where the 
components are uncorrelated. The parameter $s$ has been equal to 15.
We do not consider the nuclear norm of the matrix as defined in 
(\ref{N2SID}), but instead we first multiply it with a random matrix
from the right. This random matrix was obtained from a standardized normal 
density function and the number of columns was 22. It is known 
that this type of randomization is a powerful tool in low rank
matrix approximation, \cite{hal+mar+tro11}. The reason we do this modification
is that it will reduce the computational complexity without significantly
affecting the quality of the results obtained. We choose model order by 
looking at the singular values of the matrix that minimizes the nuclear norm. 
The model order is equal to the index of the singular values that is 
closest to the logarithmic mean of the largest and smallest singular value. 
However, in case that index is greater than 10 we choose as model order 10. 

Since our approach gives
as solution a whole family of models parameterized with regularization 
parameter $\lambda$ we perform a second optimization over this parameter with
respect to a fit criterion as implemented in the command {\tt FIT} of 
the System Identification Toolbox in Matlab. We consider 20 logarithmically 
spaced values of $\lambda/N$ in the interval $[10^{-0.5},10^4]$. 

\begin{figure}
\includegraphics[width=.8\linewidth]{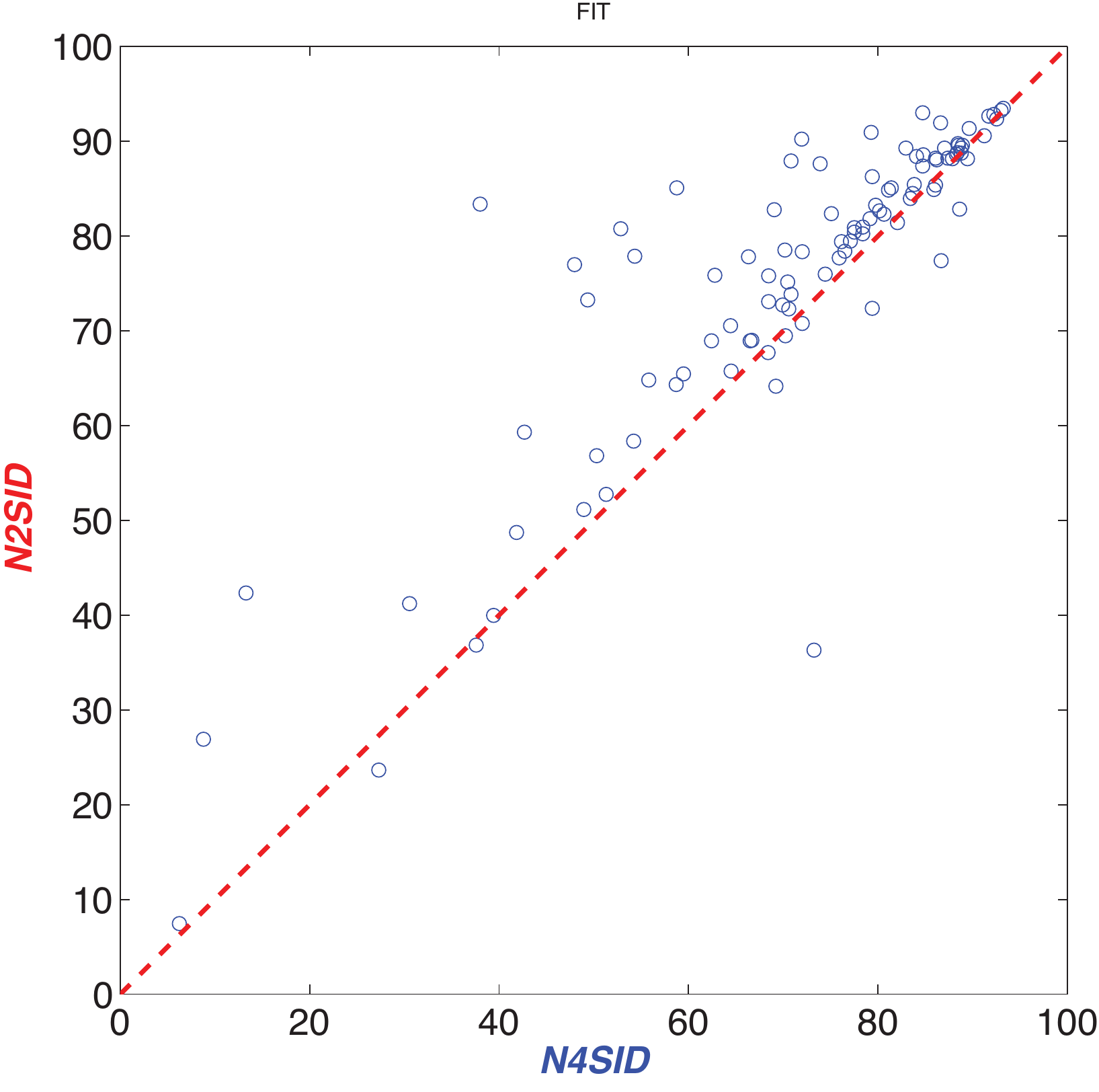}
\caption{Scatter plot showing the fit N2SID versus N4SID for 100 
randomly generated examples.}
\label{fig:scatter}
\end{figure}

We then compare our method with a standard subspace method as implemented in 
the command {\tt n4sid} of the System Identification Toolbox of Matlab. 
This algorithm will pick model order equal to the one in the 
interval $0$ to $10$ which give the best fit on the identification data.
We also make sure that {\tt n4sid} uses the same value of $s$ as we do. We
also insist that the direct term in the model is estimated.  
The comparison between our model and the one of {\tt n4sid} is done by 
computing the fit of the two models on a second validation data 
set. In Figure \ref{fig:scatter} the fit of N2SID versus N4SID is 
plotted for the 100 random examples. 
In this plot one negative fit value has been removed. In more than 80\% of
the cases N2SID has a higher fit value than N4SID. The average fit values
are 76.0 for N2SID and 71.1 for N4SID. The average model orders are 
6.67 for N2SID and 5.01 for N4SID. This clearly demonstrates the advantage
of using N2SID on short data sets. We have also performed experiments
on longer data sets. Preliminary results show that there is no significant 
difference between the two methods. This is an indication that for
large data length batches both methods deliver consistent results. 

\subsection{Closed-loop experiment}

\begin{figure}
\begin{center}
\includegraphics[width=.75\linewidth]{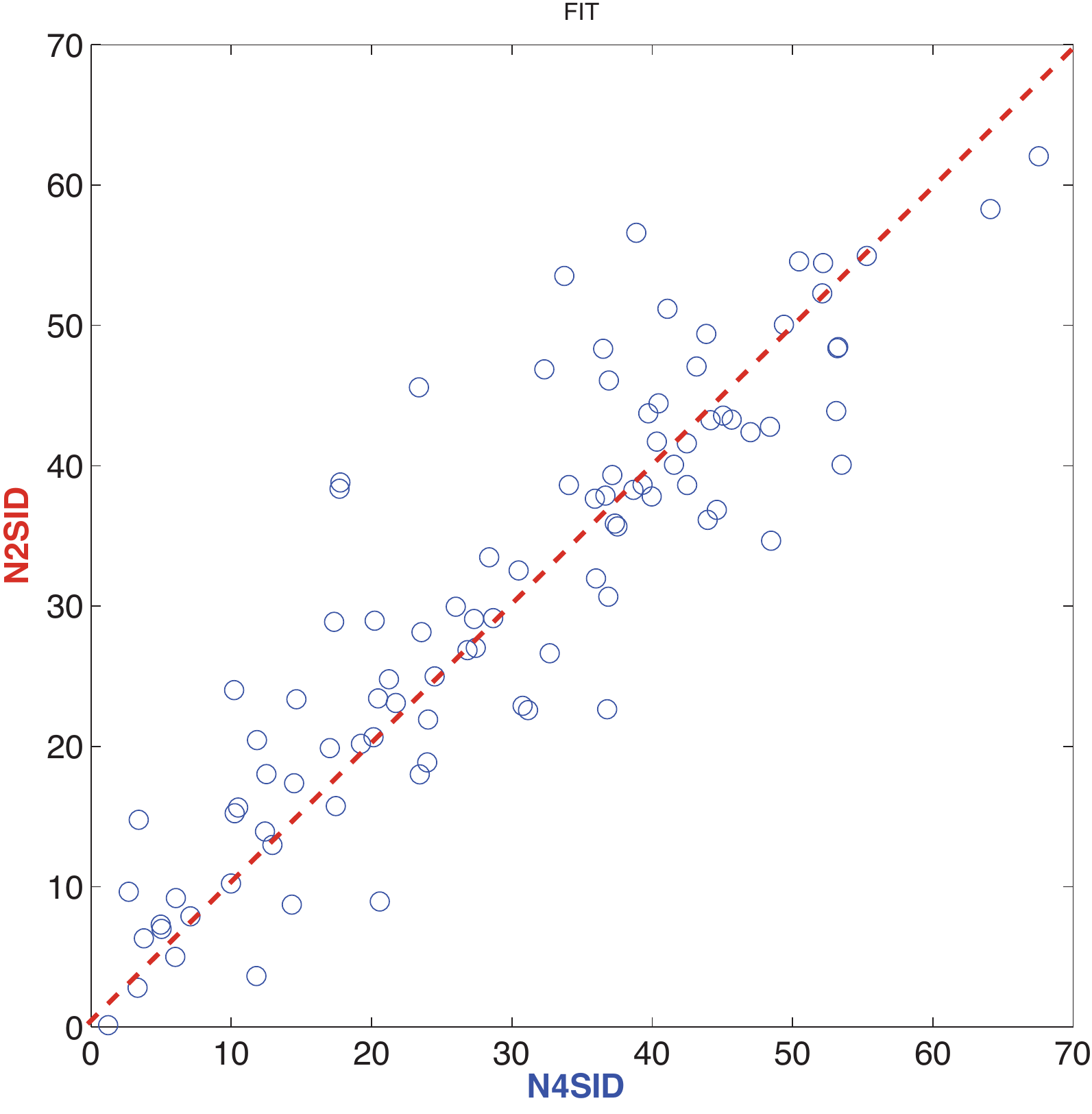}
\caption{Scatter plot showing the fit N2SID versus N4SID for 100 
different realizations of the reference value and the noise.}
\label{fig:fit_scatter_closed}
\end{center}
\end{figure}

\begin{figure}
\begin{center}
\hspace*{-1cm}\includegraphics[width=1.2\linewidth]{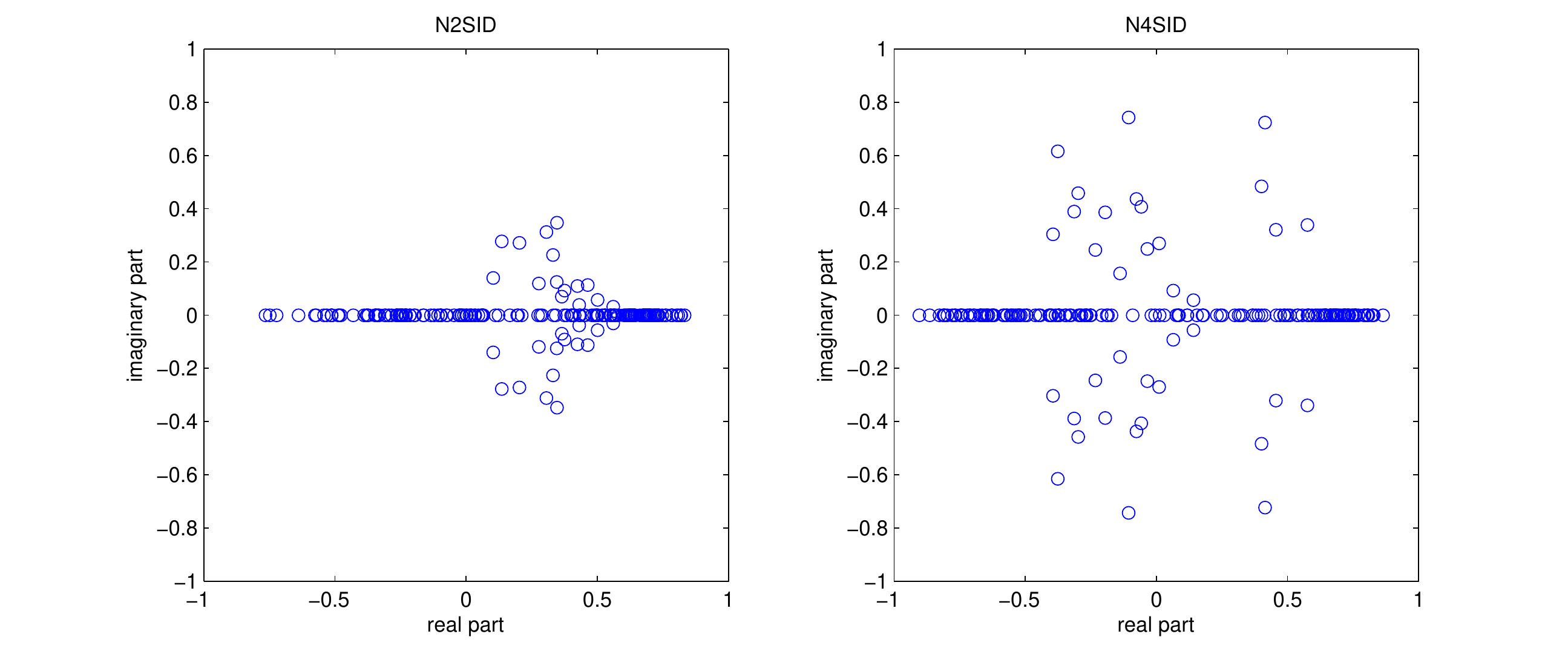}
\caption{Scatter plot showing the eigenvalues of the 
$A$-matrix for N2SID and N4SID for 100 
different realizations of the reference value and the noise.}
\label{fig:eig_scatter_closed}
\end{center}
\end{figure}

In the second example we consider closed loop identification. For this 
purpose we consider a model on the form (\ref{ssmI}), where 
\begin{align*}
A&=\begin{bmatrix}0&1\\0&0.7\end{bmatrix};\quad
B=\begin{bmatrix}0\\1\end{bmatrix};\quad
K=\begin{bmatrix}-0.3\\0.04\end{bmatrix}\\
C&=\begin{bmatrix}1&0\end{bmatrix};\quad D = 0
\end{align*}
We control the model with an observer-based state feedback where $K$ is the
observer gain and where the state feedback matrix is 
$L=\begin{bmatrix}0.25&-0.3\end{bmatrix}$. This design will place the closed
loop eigenvalues all at 0.5. The controller also has a feed-forward from a 
reference value $r(k)$, which is a sequence of 
$\pm 1$ obtained by taking the sign of a vector of values obtained from 
a standardized normal density function. This signal is multiplied with 
a gain before it is added to the input signal such that the closed loop 
steady gain from reference value to $y(k)$ is equal to one. The noise 
$e(k)$ is white and drawn from a normal density function with standard 
deviation equal to 0.1. In this experiment we will not look for the best
model order, but instead we just compute models of order $n=2$. All the 
other parameters are the same as in the previous experiment. We show in
Figure \ref{fig:fit_scatter_closed} the fit of N2SID versus N4SID 
for 100 different realizations of the reference value and the noise. 
The fit is better for N2SID in 59 \% of the cases. Moreover, as is seen 
in Figure \ref{fig:eig_scatter_closed}, the spread of the eigenvalues of 
the $A$-matrix is smaller for N2SID as compared to N4SID. 
\section{Concluding Remarks}

Subspace identification is revisited in this paper in the scope of
nuclear norm optimization methods. A new way to impose structural
matrix properties in the data equation in subspace identification on
the measured data has been presented. The new subspace identification
method is referred to as N2SID. It is shown that especially for small
data length batches when the number of samples is only a small
multiple of the order of the underlying system, the incorporation of
structural information about the low rank property of the matrix
revealing the required subspace and the (block) Toeplitz structure of
the matrix containing the unknown Markov parameters enables to improve
the results of widely used SID methods. In addition to the structural
constraints the N2SID method also enables to make a trade-off in the
first step of the calculations 
between the subspace approximation and the prediction error
cost function. As such it also overcomes the persistent drawback that
SID did not consider a (classical prediction error) cost function.

The single integrative step that aims at imposing key structural matrix 
properties makes a trade-off between the prediction error cost function and 
the problem to retrieve the subspace of interest. This integrative approach 
may help to simplify the analysis of the optimality of SID methods and to 
further clarify their link with prediction error methods.
This new way of looking upon SID will open up the possibility for new 
developments in the future.

\bibliography{refN2SID,cvx-refs2,database2,myref,subspace}
\end{document}